\documentclass{elsart}

\journal{Optics Communications}
\date{received 23 April 2002}         
\newcommand{\I}{\mathrm{i}}
\newcommand{\Exp}[1]{{\mathrm{e}}^{\mbox{\footnotesize$#1$}}}
\newcommand{\power}[1]{^{\mbox{\footnotesize$#1$}}}
\newcommand{\rewop}[1]{_{\mbox{\footnotesize$#1$}}}
\newcommand{\tfrac}[2]{{\textstyle\frac{#1}{#2}}}
\newcommand{\adj}{^{\dagger}}
\newcommand{\thalf}{\tfrac{1}{2}}
\newcommand{\pra}{Phys.\ Rev.\ A\ }
\newcommand{\prl}{Phys.\ Rev.\ Lett.\ }
\newcommand{\oc}{Opt.\ Commun.\ }

\begin{document}

\begin{frontmatter}

\title{Statistics of dressed modes in a thermal state}

\author[MATH,PHYS]{Berthold-Georg Englert},
\author[MATH,PHYS]{Stephen A. Fulling},
\author[PHYS,Berk]{Mark D. Pilloff}
\address[MATH]{Department of Mathematics, Texas A\&M University,\\ 
College Station, Texas 77843-3368, USA}
\address[PHYS]{Institute for Quantum Studies 
and Department of Physics,\\ 
Texas A\,\&\,M University, College Station, TX 77843-4242, USA}
\address[Berk]{Department of Physics, University of California, \\
Berkeley, California, 94720, USA}

\begin{abstract}%
By a Wigner-function calculation,
we evaluate the trace of a certain Gaussian 
operator arising in the theory of a boson system subject to both 
finite temperature and (weak) interaction.
Thereby we rederive (and generalize) a recent result by
Kocharovsky, Kocharovsky, and Scully 
[\pra \textbf{61}, 053606 (2000)] 
in a way that is technically much simpler.
One step uses a special case of the response 
of Wigner functions to linear transformations, 
and we demonstrate the general case by simple means.
As an application we extract the counting statistics for 
each mode of the Bose gas.
\end{abstract}
\end{frontmatter}

In a recent paper on dilute boson gases \cite{KKS:00}, 
Kocharovsky, Kocharovsky, and Scully faced the 
problem of evaluating the trace of a product of two 2-mode Gaussian 
operators. 
They solved it by exploiting the algebra of 2-mode squeezing operators
in a remarkable \emph{tour de force} occupying the appendix of \cite{KKS:00}. 
The main objective of the present Brief Report is to put on record an 
alternative derivation that is simpler and more direct. 

In a few words, the essence of our calculation is this: We recall 
(i) that the trace of a product of two operators can be obtained as 
a phase space integral of the product of the respective Wigner 
functions \cite{fn:1}; (ii) that the Wigner functions of Gaussian 
operator functions are Gaussian functions of the phase space 
variables; and (iii) that integrals of Gaussians are immediate. 

The trace of interest is
\begin{equation}
  \label{eq:A1}
  T(u,v;\epsilon)=\mathrm{tr}\bigl\{G(u,v)\rho_{\epsilon}\bigr\}
=\bigl\langle G(u,v)\bigr\rangle_{\epsilon}
\end{equation}
with 
\begin{equation}
  \label{eq:A2}
  G(u,v)=\Exp{\I ua\adj a+\I vb\adj b}
\end{equation}
and
\begin{equation}
  \label{eq:A3}
  \rho_{\epsilon}=\bigl(1-\Exp{-\epsilon}\bigr)^2
                  \Exp{-\epsilon(A\adj A+B\adj B)}\,,
\end{equation}
where $a\adj,a$ and $b\adj,b$ are the ladder operators of the physical
excitations (two independent harmonic oscillators), and the Bogolubov
transformation
\begin{equation}
  \label{eq:A4}
  \begin{array}[b]{rcl@{\quad}rcl}
A&=&a\cosh\vartheta-b\adj\sinh\vartheta\,,&
A\adj&=&a\adj\cosh\vartheta-b\sinh\vartheta\,,\\
B&=&b\cosh\vartheta-a\adj\sinh\vartheta\,,&
B\adj&=&b\adj\cosh\vartheta-a\sinh\vartheta
  \end{array}
\end{equation}
(with real $\vartheta$) relates them to the ladder operators of the 
``dressed modes.''
The statistical operator $\rho_{\epsilon}$ is a thermal state 
(with temperature parameter $\epsilon>0$) that is diagonal
in the dressed number operators $A\adj A$ and $B\adj B$.
Table~\ref{tbl:1} states how our notation is related to the one used in
\cite{KKS:00}. 
The factor $G(u,v)$ is the generating function for the count of physical
excitations inasmuch as
\begin{equation}
  \label{eq:A5}
  \delta\rewop{a\adj a,m}\delta\rewop{b\adj b,n}
=\int\limits_{(2\pi)}\frac{\d u}{2\pi}\int\limits_{(2\pi)}\frac{\d v}{2\pi}
\,\Exp{-\I mu-\I nv}G(u,v)\,,
\end{equation}
the integrations covering any interval of length $2\pi$.
Accordingly, upon expanding $T(u,v;\epsilon)$ in powers of $\Exp{\I u}$ and
$\Exp{\I v}$ we get the probabilities for having certain numbers of quanta in
the physical excitations.

\begin{table}[!t]
\caption{\label{tbl:1}%
Dictionary for switching from the present notation to the
one in Eqs.~(63), (68), (69) of Ref.~\cite{KKS:00}.}
\begin{center}
\parbox{0.3\columnwidth}{%
\begin{tabular}{cc}\hline\hline  
here & there \\ \hline
$T(u,u;\epsilon)$ & $\Theta_{\pm\mathbf{k}}(u)$\\
$a,a\adj$ & $\hat{\beta}_{\mathbf{k}},\hat{\beta}_{\mathbf{k}}\adj$\\
$b,b\adj$ & $\hat{\beta}_{-\mathbf{k}},\hat{\beta}_{-\mathbf{k}}\adj$\\
$A,A\adj$ & $\hat{b}_{\mathbf{k}},\hat{b}_{\mathbf{k}}\adj$\\
$B,B\adj$ & $\hat{b}_{-\mathbf{k}},\hat{b}_{-\mathbf{k}}\adj$\\
\hline\hline  
\end{tabular}
}\quad
\parbox{0.3\columnwidth}{%
\begin{tabular}{cc} \hline\hline  
here & there \\ \hline
$u,v$ & $u,u$ \\
$\epsilon$ & $\varepsilon_{\mathbf{k}}/T$\\
$\tanh\vartheta$ & $A_{\mathbf{k}}$\\
$Y_+$ & $z(-A_{\mathbf{k}})$\\
$Y_-$ & $z(A_{\mathbf{k}})$\\
\hline\hline
\end{tabular}
}
\end{center}
\end{table}

\section{Phase-space evaluation}\label{sec:TheTRACE}
Our starting point is the 1-mode thermal state,
\begin{equation}
\rho_{\epsilon}^{(1)}=(1-\Exp{-\epsilon})\Exp{-\epsilon a\adj a}\,,
\label{eq:A6}  
\end{equation}
and its well known Wigner function,
\begin{equation}
\bigl[\rho_{\epsilon}^{(1)}\bigr]_{\mathrm{W}}=2\tanh(\thalf\epsilon)
\Exp{-2\alpha^*\alpha\tanh(\half\epsilon)}\,,
\label{eq:A7}  
\end{equation}
normalized in accordance with
\begin{equation}
\mathrm{tr}\bigl\{\rho_{\epsilon}^{(1)}\bigr\}
=\int\frac{\d^2\alpha}{\pi}\,\bigl[\rho_{\epsilon}^{(1)}\bigr]_{\mathrm{W}}\,,
\label{eq:A8}
\end{equation}
where $\alpha$ is the complex phase space variable associated with $a$, and
the standard parameterization of the phase space integral 
--- specified by $\alpha=x+\I y$, $\alpha^*=x-\I y$, 
and $\d^2\alpha=\d x\,\d y$ --- suffices for our purposes.

First, we need the Wigner function (or Weyl transform) of $G(u,v)$ 
in (\ref{eq:A2}), which (among other methods) can be found efficiently by
analytic continuation from (\ref{eq:A6}) and (\ref{eq:A7})  
[taking $\epsilon\to-\I u$ and $\epsilon\to-\I v$ in two copies of 
(\ref{eq:A7}), discarding its prefactor]:
\begin{equation}
\begin{array}[b]{rcl}
\bigl[G(u,v)\bigr]_{\mathrm{W}}&=&
\Exp{-\half\I(u+v)}\sec(\thalf u)\sec(\thalf v)
 \\&&\times
\Exp{2\I\alpha^*\alpha\tan(\half u)+2\I\beta^*\beta\tan(\half v)}\,,
\end{array}
\label{eq:A9}
\end{equation}
where $\beta$ is the phase space variable for $b$.

Second, to get the Wigner function of the statistical operator 
(\ref{eq:A3}) we recall a fundamental property of Wigner functions, 
namely that linear similarity transformations of the ladder 
operators carry over to the Wigner function 
(see Sec.~\ref{sec:SimTrafo} below).
For the Bogolubov transformation (\ref{eq:A4}) this means that 
$\bigl[F(a\adj,a,b\adj,b)\bigr]_{\mathrm{W}}
 =f(\alpha^*,\alpha,\beta^*,\beta)$
implies \cite{fn:2}
\begin{eqnarray}
\bigl[F(A\adj,A,B\adj,B)\bigr]_{\mathrm{W}}
=f(\alpha^*\cosh\vartheta-\beta\sinh\vartheta,\dots,
   \beta\cosh\vartheta-\alpha^*\sinh\vartheta)\,,&&
\nonumber\\
  \label{eq:A10}
\end{eqnarray}
and therefore we have
\begin{eqnarray}
    \bigl[\rho_{\epsilon}\bigr]_{\mathrm{W}}&=&4\tanh^2(\thalf\epsilon)
\nonumber\\ &&\times
\Exp{-2[(\alpha^*\alpha+\beta^*\beta)\cosh(2\vartheta)
    -(\alpha^*\beta^*+\alpha\beta)\sinh(2\vartheta)]\tanh(\half\epsilon)}\,.
\label{eq:A11}
\end{eqnarray}

Given (\ref{eq:A9}) and (\ref{eq:A11}), 
the trace (\ref{eq:A1}) is a 4-dimensional Gaussian integral that is 
routinely evaluated,
\begin{equation}
\begin{array}[b]{rcl}
  T(u,v;\epsilon)&=&\displaystyle
\int\frac{\d^2\alpha}{\pi}\int\frac{\d^2\beta}{\pi}\,
                  \bigl[G(u,v)\bigr]_{\mathrm{W}}\, 
                  \bigl[\rho_{\epsilon}\bigr]_{\mathrm{W}}
  \\
&=& \displaystyle
4(\Exp\epsilon-1)^2 \Bigl[ (\Exp\epsilon-1)^2 (1+\Exp{\I u})(1+\Exp{\I v}) 
  \\ &&\hphantom{4(\Exp\epsilon-1)^2 \Bigl[}+ 
(\Exp\epsilon+1)^2 (1-\Exp{\I u})(1-\Exp{\I v})
  \\ &&\hphantom{4(\Exp\epsilon-1)^2 \Bigl[}+ 
2(\Exp{2\epsilon}-1)(1-\Exp{\I u+\I v})\cosh(2\vartheta)\Bigr]^{-1}\,.
\end{array}
 \label{eq:A12primitive}
\end{equation}
While this formula is the most direct and 
natural result of the calculation, it may appear physically 
inscrutable. 
However, it can be worked into the form   
\begin{equation}
\begin{array}[b]{rcl}
  T(u,v;\epsilon) 
 &=&\displaystyle
\frac{2(Y_+-1)(Y_--1)}
      {(Y_+-\Exp{\I u})(Y_--\Exp{\I v})+(Y_--\Exp{\I u})(Y_+-\Exp{\I v})}
 \\
&=&T(v,u;\epsilon)
\,,
\end{array}
  \label{eq:A12}
\end{equation}
where 
\begin{equation}
  \label{eq:A13}
  Y_+=\frac{\Exp{\epsilon}+\tanh\vartheta}{1+\Exp{\epsilon}\tanh\vartheta}
\,,\qquad
  Y_-=\frac{\Exp{\epsilon}-\tanh\vartheta}{1-\Exp{\epsilon}\tanh\vartheta}\,.
\end{equation}
This matches the result, for the special case $u=v$,
reported in (69) of \cite{KKS:00} 
(translated with Table~\ref{tbl:1}):
\begin{equation}
  \label{eq:A14}
  T(u,u;\epsilon)=\frac{Y_+-1}{Y_+-\Exp{\I u}}\frac{Y_--1}{Y_--\Exp{\I u}}
   =\bigl\langle\Exp{\I u(a\adj a+b\adj b)}\bigr\rangle_{\epsilon}\,.
\end{equation}
The challenge of rederiving (\ref{eq:A14}) by simpler means has 
thus been met; in fact, 
(\ref{eq:A12}) is  more general because $u$ and $v$ need not be the 
same.

\emph{Remark:}   
One can also calculate the trace (\ref{eq:A1}) by using the Q-function for 
$\rho_\epsilon$ and the P-function for $G(u,v)$, rather than the Wigner
function for both. 
But that approach is more involved because the statement corresponding to
(\ref{eq:A10}) is more complicated.

\section{Occupation probabilities}\label{sec:CountQuanta}
Let us now see what we can learn from this gain in flexibility.
Special cases are those of $v=0$ and $u=0$, the generating functions for the
count of excitations in either the $a$ mode or the $b$ mode,
\begin{equation}
\begin{array}[b]{rcl}
  T(u,0;\epsilon)&=&\displaystyle
\bigl\langle\Exp{\I ua\adj a}\bigr\rangle_{\epsilon}
                  =\frac{Y-1}{Y-\Exp{\I u}}
\,,
 \\[1.5ex]
  T(0,v;\epsilon)&=&\displaystyle
\bigl\langle\Exp{\I vb\adj b}\bigr\rangle_{\epsilon}
                  =\frac{Y-1}{Y-\Exp{\I v}}
\,,
\end{array}
  \label{eq:B1}
\end{equation}
where
\begin{equation}
  \label{eq:B2}
  Y=\frac{2Y_+Y_--Y_+-Y_-}{Y_++Y_--2}
   =\frac{\Exp{\epsilon}+\tanh^2\vartheta}{1+\Exp{\epsilon}\tanh^2\vartheta}
   >1\,.
\end{equation}
The symmetry in (\ref{eq:B1}) is, of course, just 
a particular example of
$T(u,v;\epsilon)=T(v,u;\epsilon)$, which we noted at (\ref{eq:A12}) in passing.
The geometric series that result from expanding the right-hand sides of
(\ref{eq:B1}) in powers of $\Exp{\I u}$ or $\Exp{\I v}$, respectively, tell us the
individual counting statistics for each mode,
\begin{equation}
  \label{eq:B3}
  \bigl\langle\delta\rewop{a\adj a,n}\bigr\rangle_{\epsilon}=
  \bigl\langle\delta\rewop{b\adj b,n}\bigr\rangle_{\epsilon}=
  (1-Y^{-1})Y^{-n}\,.
\end{equation}
So, if there were no statistical correlations between the two modes, the
probability for having a total of $N$ quanta in both modes would be given by
\begin{equation}
  \label{eq:B4}
  \sum_{n=0}^N\bigl\langle\delta\rewop{a\adj a,n}\bigr\rangle_{\epsilon}
              \bigl\langle\delta\rewop{b\adj b,N-n}\bigr\rangle_{\epsilon}
   =(N+1) (1-Y^{-1})^2Y^{-N}\,.
\end{equation}
But there \emph{are} correlations and, indeed, the correct value of
$\bigl\langle\delta\rewop{a\adj a+b\adj b,N}\bigr\rangle_{\epsilon}$, obtained
by expanding $T(u,u;\epsilon)$ of (\ref{eq:A14}) in powers of $\Exp{\I u}$,
\begin{equation}
  \label{eq:B5}
 \bigl\langle\delta\rewop{a\adj a+b\adj b,N}\bigr\rangle_{\epsilon}
= (1-Y_+^{-1})(1-Y_-^{-1})
  \frac{Y_+^{-N-1}-Y_-^{-N-1}}{Y_+^{-1}-Y_-^{-1}}\,,
\end{equation}
differs noticeably from (\ref{eq:B4}), unless we are in the limiting situation
of $\vartheta=0$ when $Y_+=Y_-=Y$. 
In that limit we have
\begin{equation}
  \label{eq:B6}
  T(u,v;\epsilon)\bigl|\rewop{\vartheta=0}
  =\frac{\Exp{\epsilon}-1}{\Exp{\epsilon}-\Exp{\I u}}
   \frac{\Exp{\epsilon}-1}{\Exp{\epsilon}-\Exp{\I v}}\,, 
\end{equation}
and this factorization states the obvious:
For $\vartheta=0$ the modes do not get dressed and so they maintain their
statistical independence.

Rather more involved is the expression for the probability of $m$ quanta in
mode $a$ and $n$ quanta in mode~$b$,
\begin{equation}
\begin{array}[b]{rcl}
\bigl\langle\delta\rewop{a\adj a,m}
             \delta\rewop{b\adj b,n}\bigr\rangle_{\epsilon}
&=&\displaystyle
(1-Y_+^{-1})(1-Y_-^{-1})
\bigl[\thalf(Y_+^{-1}+Y_-^{-1})\bigr]^{m+n}  
 \\ &&\displaystyle \times
\sum_{k=0}^{\infty}\frac{(m+n-k)!}{k!(m-k)!(n-k)!}
\left[\frac{-4Y_+Y_-}{(Y_++Y_-)^2}\right]^k\,,  
\end{array}
\label{eq:B7}
\end{equation}
where the summation terminates when $k$ equals the smaller one of $m$ and $n$.
Note that the right-hand side is invariant under the interchange
$m\leftrightarrow n$, which is another manifestation of
$T(u,v;\epsilon)=T(v,u;\epsilon)$. 
Incidentally, one could write (\ref{eq:B7}) in terms of a Jacobi polynomial,
but that doesn't seem to add transparency to the result.

\section{Bogolubov transformation of a single mode}\label{sec:1mode}
Also worthy of study is a Bogolubov transformation mixing one mode with 
itself by
\begin{equation}
A = a \cosh\vartheta - a\adj\sinh\vartheta
\label{eq:1modebogo}
\end{equation}
in place of (\ref{eq:A4}).
In fact, the dressing of a pair of modes as in (\ref{eq:A4}) can be 
diagonalized into two decoupled transformations of the form 
(\ref{eq:1modebogo}); in the Bose condensation problem this corresponds 
to replacing traveling waves by standing waves.

In this case we should evaluate
\begin{equation}
T(u;\epsilon)= \int{\d^2\alpha\over \pi} 
\,\bigl[G(u)\bigr]_W \,\bigl[\rho_\epsilon\bigr]_W\,,
\label{eq:1modetr}
\end{equation}
where $\bigl[G(u)\bigr]_W$ is the $v=0$ version of
(\ref{eq:A9}) 
and
\begin{equation}
\label{eq:1moderho}
\bigl[\rho_\epsilon\bigr]_W =
2 \tanh(\thalf\epsilon)
\Exp{-[2\alpha^*\alpha \cosh(2\vartheta) - (\alpha^{*2}+\alpha^2)
\sinh(2\vartheta)]\tanh(\half\epsilon)} 
\end{equation}
is obtained by $\alpha\to\alpha\cosh\vartheta-\alpha^*\sinh\vartheta$ 
in (\ref{eq:A7}).
[Note the subtle difference in the relative sizes of the terms in the 
exponent in (\ref{eq:1moderho}) versus (\ref{eq:A11}).]
The Gaussian integration is more elementary than in the previous 
case and yields
\begin{equation}
\begin{array}[b]{rcl}
T(u;\epsilon)&=& \displaystyle
2(\Exp{\epsilon}-1)\Bigl[
 (\Exp{\epsilon}-1)^2(1+\Exp{\I u})^2 
+(\Exp{\epsilon}+1)^2(1-\Exp{\I u})^2 
 \\ && \hphantom{2(\Exp{\epsilon}-1)\Bigl[}
 +2(\Exp{2\epsilon}-1)(1-\Exp{2\I u})\cosh(2\vartheta)\Bigr]^{-\frac{1}{2}}
\end{array}
\label{eq:1modeT} 
\end{equation}
as the analog of (\ref{eq:A12primitive}).
Note that $T(u;\epsilon)^2 =T(u,u;\epsilon)$, as it should.
On the other hand, $T(u;\epsilon)T(v;\epsilon)$ does not equal
$T(u,v;\epsilon)$, as these functions are giving the statistics 
for two different bases of single-quantum states, which have different 
dynamics.
If desired, the counterpart of (\ref{eq:B7}) could be obtained by expanding 
(\ref{eq:1modeT}) in powers of $\Exp{\I u}$, which is easily done because we
meet here a familiar generating function for Legendre's polynomials.   

One motivation for investigating a single mode was the hope of 
elucidating the structure of equations (\ref{eq:A13}), which cry out to 
be interpreted as addition formulas for the hyperbolic tangent function 
with $\Exp{\epsilon} = \tanh (\ldots)$.
There is, in fact, a way of constructing the thermal state of a boson 
mode by means of another Bogolubov transformation \cite{UT:75};
one would think that (\ref{eq:A13}) then expresses the composition of the 
two transformations. 
However, the appropriate thermal parameter in that construction
is $\thalf\epsilon$ rather than $\epsilon$.  This discrepancy may be 
related to the fact, stressed in \cite{KKS:00}, that quanta in the 
interacting Bose gas are present in strongly correlated \emph{pairs}.
 Calculations following \cite{UT:75} are in progress, but so far have not 
yielded expressions as simple as those provided by the method reported 
here.

\section{Linear similarity transformations}\label{sec:SimTrafo}
It is striking that the fundamental transformation property  
of Wigner functions, 
of which (\ref{eq:A10}) is the special case we make use of, 
is not mentioned in any of the  
reviews \cite{Tatarskii:83,BJ:84,HOSW:84} or textbooks
\cite{SZ:97,Schleich:01}, although it has been much exploited for
various applications (e.g., in \cite{Englert:89,newRefs}).
Since it deserves to be known more widely, we present the following
compact derivation, which is more general than all previous ones.

As a warm-up we'll first consider the single-mode case of Sec.~\ref{sec:1mode}.
The Wigner function (or Weyl transform) $F_{\mathrm{W}}(\alpha^*,\alpha)$ of
an operator function $F(a\adj,a)$ is given by
\begin{equation}
\begin{array}[b]{rcl}
F_{\mathrm{W}}(\alpha^*,\alpha)&=&\displaystyle
\int\frac{\d^2\beta}{\pi}\,
\Exp{\beta\alpha^*-\beta^*\alpha}\mathrm{tr}\left\{
\Exp{\beta^*a-\beta a\adj}F(a\adj,a)\right\} \\[2ex]
&=&2\,\mathrm{tr}\left\{
\Exp{\alpha a\adj-\alpha^* a}S\Exp{\alpha^* a-\alpha a\adj}F(a\adj,a)
\right\}
\end{array}
  \label{eq:C1}
\end{equation}
where we encounter the unitary displacement operator 
$\Exp{\alpha^* a-\alpha a\adj}$,
\begin{equation}
\begin{array}[b]{rcl}
\Exp{\alpha a\adj-\alpha^* a}\,a\adj\,\Exp{\alpha^* a-\alpha a\adj}
&=&a\adj-\alpha^*\,, \\  
\Exp{\alpha a\adj-\alpha^* a}\,a\,\Exp{\alpha^* a-\alpha a\adj}
&=&a-\alpha\,,
\end{array}
  \label{eq:C2}
\end{equation}
and the reflection operator 
\begin{equation}
  \label{eq:C3}
  S=\int\frac{\d^2\beta}{2\pi}\,\Exp{\beta^*a-\beta a\adj}\,,
\end{equation}
which is both unitary and Hermitean and therefore its own inverse,
$S\adj= S=S^{-1}$ \cite{fn:3}. 
Its name derives from the unitary transformation that $S$ effects,
\begin{equation}
  \label{eq:C4}
  S^{-1}a\adj S=-a\adj\,,\qquad S^{-1}aS=-a\,.
\end{equation}

A linear similarity transformation turns $F(a\adj,a)$ into $G(a\adj,a)$,
another operator function,
\begin{equation}
  \label{eq:C5}
  G(a\adj,a)=V^{-1}F(a\adj,a)V=F(\mu a\adj+\nu a,\sigma a+\tau a\adj)\,,
\end{equation}
where
\begin{equation}
  \label{eq:C6}
  \begin{array}[b]{rcl@{\qquad}rcl}
V^{-1}a\adj V&=&\mu a\adj+\nu a\,, &V a\adj V^{-1}&=&\sigma a\adj -\nu a\,,\\
V^{-1}a V&=&\sigma a +\tau a\adj\,, &V a V^{-1}&=&\mu a -\tau a\adj\,.
  \end{array}
\end{equation}
The commutation relation $[a,a\adj]=1$ puts the restriction
$\mu\sigma-\nu\tau=1$ on the numerical coefficients $\mu,\nu,\sigma,\tau$, 
and otherwise they can take on any complex values permitted by
$|\mu|,|\sigma|\geq|\nu|,|\tau|$, which condition ensures that 
$V^{-1}a\adj V$ has eigenbras and $V^{-1}a V$ has eigenkets.
It is not necessary that (\ref{eq:C6}) be a unitary transformation, such as
(\ref{eq:1modebogo}). 
In particular, the non-unitary transformation $a\adj\to a\adj-a$,
$a\to\frac{1}{2}(a+a\adj)$ is permissible and useful \cite{Englert:89}.

The crucial observation is now that the reflection operator is invariant under
this similarity transformation,
\begin{equation}
  \label{eq:C7}
  V^{-1}SV=S\qquad\mbox{or}\qquad S^{-1}VS=V\,.
\end{equation}
This is so because $V$ is the exponential of a linear combination of
${a\adj}^2$, $a\adj a$, and $a^2$ (or a product of such exponentials), as
illustrated by 
\begin{equation}
  \label{eq:C7'}
V=\Exp{\half(\tau/\mu){a\adj}^2}
  \mu\power{-a\adj a}\Exp{-\half(\nu/\mu)a^2}\,.  
\end{equation}
Clearly, the reflection (\ref{eq:C4}) has no effect on $V$, which is the
second statement of (\ref{eq:C7}).

Now, perform the replacements 
\begin{equation}\begin{array}[b]{rcl}
F(a\adj,a)&\to&V^{-1}F(a\adj,a)V=G(a\adj,a)\,,\\
S&\to&S=V^{-1}SV
\end{array}  \label{eq:C9}
\end{equation}
in (\ref{eq:C1}) to establish
\begin{equation}
  \label{eq:C10}
  G_{\mathrm{W}}(\alpha^*,\alpha)=2\,\mathrm{tr}\left\{
V\Exp{\alpha a\adj-\alpha^* a}V^{-1}SV\Exp{\alpha^* a-\alpha a\adj}V^{-1}
F(a\adj,a)\right\}\,,
\end{equation}
then note that
\begin{equation}
  \label{eq:C11}
  V\Exp{\alpha^* a-\alpha a\adj}V^{-1}=
\Exp{(\mu\alpha^*+\nu\alpha) a-(\sigma\alpha+\tau\alpha^*) a\adj}\,,
\end{equation}
and arrive at
\begin{equation}
  \label{eq:C12}
 G_{\mathrm{W}}(\alpha^*,\alpha)
= F_{\mathrm{W}}(\mu\alpha^*+\nu\alpha,\sigma\alpha+\tau\alpha^*)  \,.
\end{equation}
Indeed, the linear similarity transformation (\ref{eq:C5}) of the operator
functions carries over to their Wigner functions.

The arguments in higher dimensions are quite analogous.
There are now $n$ pairs of ladder operators and $n$ pairs of phase-space
variables associated with them. 
We write them compactly as $2n$-component rows,
\begin{equation}
\begin{array}[b]{rcl}
  \pol{a}&=&\bigl(a_1\adj,a_1,a_2\adj,a_2,\dots,a_n\adj,a_n\bigr)\,,
\\
  \pol{\alpha}&=&\bigl(\alpha_1^*,\alpha_1,\alpha_2^*,\alpha_2,\dots,
                       \alpha_n^*,\alpha_n\bigr)\,.
\end{array}  
\label{eq:C13}
\end{equation}
The $n$-dimensional displacement and reflection operators are then
\begin{equation}
  \label{eq:C14}
  \prod_{k=1}^n\Exp{\alpha_k^*a_k-\alpha_ka_k\adj}
  =\Exp{\pol{\alpha}\mathsf{K}\pol{a}^{\mathrm{T}}}
\,,\quad
  S=\int\frac{\d^{2n}\beta}{(2\pi)^n}\,
 \Exp{\pol{\beta}\mathsf{K}\pol{a}^{\mathrm{T}}}
\end{equation}
with the $2n\times2n$ block matrix
\begin{equation}
  \label{eq:C16}
  \mathsf{K}=\left(
    \begin{array}{rrcrr}
           0&-1& &&\\
           1& 0& &&\\
           &&\ddots &&\\
           &&& 0&-1\\ 
           &&& 1& 0
    \end{array}
\right)=-{\mathsf{K}}^{-1}=-{\mathsf{K}}^{\mathrm{T}}\,,
\end{equation}
so that the analog of (\ref{eq:C1}) reads
\begin{equation}
  \label{eq:C17}
  F_{\mathrm{W}}(\pol{\alpha})=2^n\,\mathrm{tr}\left\{
\Exp{-\pol{\alpha}\mathsf{K}\pol{a}^{\mathrm{T}}}S
\Exp{\pol{\alpha}\mathsf{K}\pol{a}^{\mathrm{T}}}F(\pol{a})\right\}\,.
\end{equation}
A linear similarity transformation in $n$ dimensions is of the form
\begin{equation}
  \label{eq:C18}
V^{-1}\pol{a}\,V=\pol{a}\,\mathsf{V}\,,\qquad  
V\pol{a}\,V^{-1}=\pol{a}\,\mathsf{V}^{-1}
\end{equation}
where $\mathsf{V}$ is a $2n\times2n$ matrix restricted mainly by
\begin{equation}
  \label{eq:C19}
  \mathsf{V}\mathsf{K}\mathsf{V}^{\mathrm{T}}=\mathsf{K}\,,
\rule[-10pt]{0pt}{10pt}
\end{equation}
which is the analog of $\mu\sigma-\nu\tau=1$ for the coefficients in
(\ref{eq:C6}).
Here, too, $V$ is the exponential of a bilinear form of the ladder operators,
$ V=\exp\bigl(\pol{a}\mathcal{V}\pol{a}^{\mathrm{T}}\bigr)$,
with a (largely) arbitrary complex $2n\times2n$ 
coefficient matrix $\mathcal{V}$, and
so (\ref{eq:C7}) is equally true in $n$ dimensions.
By arguments analogous to those of (\ref{eq:C9})--(\ref{eq:C12}) 
it then follows that the Wigner function of the transformed operator 
\begin{equation}
  \label{eq:C21}
  G(\pol{a})=V^{-1}F(\pol{a})V=F(\pol{a}\,\mathsf{V})
\end{equation}
is given by 
\begin{equation}
\begin{array}[b]{rcl}
 G_{\mathrm{W}}(\pol{\alpha})
&=&2^n\,\mathrm{tr}\left\{
V\Exp{-\pol{\alpha}\mathsf{K}\pol{a}^{\mathrm{T}}}V^{-1}S
V\Exp{\pol{\alpha}\mathsf{K}\pol{a}^{\mathrm{T}}}V^{-1}F(\pol{a})\right\}
\\[2ex]  
&=&2^n\,\mathrm{tr}\left\{
\Exp{-\pol{\alpha}\mathsf{VK}\pol{a}^{\mathrm{T}}}S
\Exp{\pol{\alpha}\mathsf{VK}\pol{a}^{\mathrm{T}}}F(\pol{a})\right\}
\\[2ex]  
&=&F_{\mathrm{W}}(\pol{\alpha}\,\mathsf{V})\,.
\end{array}
  \label{eq:C22}
\end{equation}
This closes the case.

\ack
We thank Marlan Scully for insisting that a simpler derivation of 
(\ref{eq:A14}) was possible and worthwhile.


\begin{thebibliography}{99}
\bibitem{fn:2}
Equation (\ref{eq:A10}) is the slightest generalization of the fact
\cite{fn:1} that the quantum dynamics of a harmonic oscillator in the Wigner
representation is identical to the classical dynamics (Liouville flow) of a
phase-space probability distribution.
\end{thebibliography}
\end{document}